# Noninvasive Focused Ultrasound Spinal Cord Stimulation in Humans: A Computational Feasibility Study

Koharu Isomura[1], Wenwei Yu[1,2], Jose Gomez-Tames[1,2]*

[1]Department of Medical Engineering, Graduate School of Engineering, Chiba University, Chiba 263-8522, Japan

[2]Center for Frontier Medical Engineering, Chiba University, Chiba 263-8522, Japan

* Corresponding author:
Jose Gomez-Tames
jgomez@chiba-u.jp

**Abstract**

**Background:** Trans-spinal FUS (tsFUS) has recently been shown promise in modulating spinal reflexes in rodents, opening new avenues for spinal cord interventions in motor control and pain management. However, anatomical differences between rodents and human spinal cords require careful targeting strategies and transducer design adaptations for human applications.
**Aim:** This study aims to computationally explore the feasibility of tsFUS in the human spinal cord by leveraging the intervertebral acoustic window and vertebral lamina.
**Method:** Acoustic simulations were performed using an anatomically detailed human spinal cord model with an adapted single-element focusing transducer (SEFT) to investigate the focality and intensity of the acoustic quantities generated within the spinal cord.
**Results:** Sonication through the intervertebral acoustic window using an adapted transducer achieved approximately 2-fold higher intensity and up to 20% greater beam overlap compared to commercial SEFT. Precise transducer positioning was critical; a 10 mm vertical shift resulted in a reduction of target intensity by approximately 7-fold. The vertebral level also substantially influenced the sonication outcomes, with the thoracic spine achieving 6-fold higher intensity than the cervical level. Sonication through the vertebral lamina resulted in approximately 2.5-fold higher intraspinal intensity in the cervical spine.
**Conclusion and Significance:** This study presents the first systematic, anatomically realistic computational model for the feasibility of tsFUS in humans. Quantifying the trade-offs between acoustic path, vertebral level, and transducer geometry offers foundational design and procedural guidelines for the design and safety of spinal FUS protocols, supporting future studies of trans-spinal FUS in humans.

# 1. Introduction

Electrical stimulation of the spinal cord is used in chronic pain management [1] and spinal cord injury rehabilitation.[2] Epidural spinal cord stimulation (eSCS) is an invasive approach in which electrodes are implanted in the epidural space to deliver controlled electrical pulses to specific areas of the spinal cord. Potential challenges with eSCS include surgical risks, the need for reoperation to replace depleted batteries, and lead migration.[4] Transcutaneous spinal cord stimulation (tSCS) is a noninvasive alternative that delivers electric currents through electrodes placed on the skin.[3] However, tSCS has limitations in targeting specific nerve areas due to its reduced focality and significant current decay, often requiring higher electric currents that can stimulate nociceptors in the skin, causing pain, and inadvertently stimulating other organs.[5]

Low-intensity focused ultrasound (FUS) is a noninvasive neuromodulation technique with superior spatial precision and deep penetration (enabling region-specific modulation). Region-specific modulation via sonication has been shown for the motor and visual areas in rabbits [6,7] and sensory-evoked potentials in humans.[8,9] Transcranial FUS (tFUS) targeting deep brain structures, such as the thalamus, elicits direct somatosensory evoked potentials.[10] Other studies have shown that tFUS applied to the frontotemporal lobe of depressed patients decreases anxiety.[11] In addition, FUS has shown promise in peripheral neuromodulation, with studies indicating the potential to modulate sensory signals in peripheral nerves to alleviate chronic pain in humans.[12] Recently, trans-spinal FUS (tsFUS) has been reported to modulate spinal reflexes in healthy rats,[13] providing evidence supporting motor control and pain management via spinal cord sonication.[14] In the rat spinal column, the limited penetration of FUS through bone tissue is less of a concern due to the relatively thin posterior arch bone, measuring approximately 6 mm.[13] In contrast, in humans, this bone is considerably thicker, exceeding 30 mm,[15] which is likely to significantly reduce FUS penetration and focality into the spinal canal. Thus, translating these findings to humans presents challenges due to interspecies anatomical differences. To bridge this gap, it is crucial to evaluate the sonication levels in humans, considering transducer parameters and spinal cord structure in a systematic manner.[16] One possibility is to use the posterior intervertebral gaps between vertebrae. Another option is to sonicate through the lamina, a flat or curved plate of the vertebra that is generally thinner than other vertebral processes.[17] Evaluating the feasibility and efficacy of these acoustic pathways is a necessary step toward optimizing transducer placement and sonication parameters for spinal applications.

Computational simulations have been fundamental to investigating the effects of cranial structure,[16] optimizing transducer parameters for targeting deep brain parts,[18] and investigating the energy distribution of low-intensity tFUS to deep brain targets.[18] However, the complex and irregular bony structure of the spine, as well as the skin-to-spine distance, presents unique characteristics compared to the skull. Additionally, shear wave transmission into the vertebrae at non-normal incident angles can affect focality intensity, and bone heating.[21] A previous computational study reported no heating risk for optimal transducer placement and highlighted significant subject-specific differences at the C5/C6 spinal level.[19] However, a systematic evaluation quantifying focality metrics such as beam overlap, along with optimization of tsFUS for both cervical and thoracic spinal levels and alternative acoustic paths, remains lacking. This is essential to provide foundational insights into the feasibility, targeting efficiency, and transducer design optimization for spinal tsFUS.

This study aims to investigate the feasibility of tsFUS through a systematic numerical analysis of ultrasound propagation characteristics in the human spinal cord. Additionally, we assess sonication through both the intervertebral acoustic window and the vertebral lamina, examining cervical and thoracic spinal levels. The analysis includes transducer geometry optimization and quantification of focality and intensity trade-offs. Our findings provide essential insights into the feasibility and design of tsFUS, supporting the emerging body of experimental research in humans.

## 2. Model and methods

*2.1. Numerical Anatomical Models and Sonication Targets*

The acoustic analysis is conducted using a detailed human anatomical model (Duke, a 34-year-old white male) from the Virtual Population database (IT'IS Foundation, Zurich, Switzerland), as shown in Figs. 1A-1B.[22] The neck and trunk regions are extracted for the simulation. Each tissue is assigned a sound velocity, mass density, and attenuation coefficient from the IT'IS database.[23] The first sonication target is the spinal cord through the posterior intervertebral acoustic window between T10-T11, selected as a suitable candidate due to its large vertebral interspace, as shown in Fig. 1C.[3] The second sonication target is a deep brain structure (thalamus), which has demonstrated neuromodulation effects in tFUS.[10,24] Acoustic levels in the thalamus provide a valuable reference for biological response thresholds when considering neuromodulation in the spinal cord. Additionally, sonication of the spinal cord along the C1 to T12 levels is further investigated (Section 3.3) through both the posterior intervertebral acoustic window and the vertebral lamina. Furthermore, the dorsal root ganglion (DRG), previously shown to be effective for chronic pain (Fig. 1C), is examined (see Appendix A).[25]

*2.2. Transducer*

Figure 2A shows the single-element focused transducer (SEFT). The reference SEFT is commercially used for brain stimulation (tFUS) and has a curvature radius ($C$) of 85 mm and an aperture width ($A$) of 80 mm.[10,24] For tsFUS, the curvature radius ($C$) and aperture width ($A$) are varied in the range of 60 to 85 mm (steps of 5 mm) and 35 to 100 mm (steps of 5 mm), respectively. The resulting number of transducers is 55, excluding those that overlap with the human trunk (Table 1). The center of the transducer is positioned to place the focal point inside the spinal cord target through either the posterior intervertebral acoustic window or the vertebral lamina.[4,16] To account for the variability of transducer positioning in normal operation, the transducer's orientation is tilted 5 and 10 degrees in four directions for a total of 9 stimulations, as shown in Fig. 2B. The frequencies investigated include 300 kHz, 500 kHz, and 700 kHz. A similar procedure was followed to position the transducer for targeting the thalamus in the brain.[24]

The sound pressure amplitude of 30 kPa is applied. It is derived from the same sonication conditions in a tFUS neuromodulation experiment,[10] where thalamic responses were elicited using a transducer delivering an $I_{SPPA}$ of 14.56 W/cm² operating in free water at 500 kHz. When the same intensity of 30 kPa is applied to the anatomical head model, the maximum intensity in the thalamus is 0.77 W/cm². This value serves as a reference neuromodulation threshold associated with confirmed biological responses.[10] This value remains below the FDA safety limit of $I_{SPPA}$ = 190 W/cm².[26] The same amplitude is used for brain and spinal cord stimulation in this study.

*2.3. Acoustic and Thermal Modeling*

The propagation of acoustic waves through heterogeneous tissues is simulated using a 3D full-wave acoustic solver in the Sim4Life platform (Zurich Med Tech, Switzerland, https://sim4life.swiss/), which encompasses all the relevant phenomena (reflections due to variations and discontinuities in the acoustic impedance, standing waves, and the combined effects of absorption and scattering).[27] Specific properties, such as the speed of sound, mass density, and attenuation coefficient, are assigned to each material at the corresponding frequency.[23] The solver uses the linear acoustic pressure wave equation (LAPWE), which has been extended and optimized for the inhomogeneous and lossy materials encountered in anatomical structures.

*2.4. Evaluation Parameters*

Maximum intensity and beam overlap are used to evaluate sonication effect and focality, respectively.[18] The spatial-peak pulse-average intensity ($I_{sppa}$) is computed from the acoustic pressure $p$ using the following equation.

$$Intensity = \frac{p^2}{2 \cdot \rho \cdot c} \times 10^{-4} \left[\frac{W}{cm^2}\right] \quad (5)$$

where $\rho$ is the mass density and $c$ is the speed of sound.
As mentioned in Section 2.2, an intensity of 0.77 W/cm² was used as the reference value for stimulating the spinal cord considering neural response in thalamic stimulation as the basis.[10] Beam overlap is defined as the percentage of the beam volume above the FWHM intensity level that is spatially confined to the target region. The FWHM (full width at half maximum) is defined as the spatial volume within which the beam intensity is at least half of its maximum value. The beam volume is illustrated in red in Fig. 2C. For instance, a 100% beam overlap means that the portion of the focal spot defined by the FWHM is fully contained within the targeted structure. The volume outside the intravertebral region (green frame in Fig. 2C) is excluded from the overlap calculation, analogous to extracranial space in brain stimulation studies where neuromodulation is not intended.

## 3. Results

*3.1. Sonication Difference in Spinal Cord and Thalamus Based on Standard SEFT*

Figure 3 shows the intensity distribution for tsFUS and tFUS at three frequencies (300 kHz, 500 kHz, and 700 kHz) using a standard SEFT size (with a radius of curvature of 85 mm and an aperture width of 80 mm). The transducer is located in the central position between T10-T11 in the case of the spinal cord. The areas in green indicate the target areas (spinal cord and thalamus). Hotspots are generated in the thalamus during tFUS, while the hotspots do not reach the spinal cord during tsFUS.

Figure 4 presents the sonication results within the target regions, as shown in Fig. 2A. Figure 4A shows that in the thalamus, the maximum acoustic intensity increases with higher frequencies. In the spinal cord, the maximum intensity is lower at 300 kHz, while there is no difference between 500 kHz and 700 kHz in maximum intensity. The stimulation intensity in the spinal cord is higher ($2.34 \pm 1.07$ times) than that in the thalamus at all frequencies. In all cases, tsFUS exceeded the neuromodulation criterion. Figure 4B shows a decrease in FWHM volume with increasing frequency in the thalamus, resulting in improved focalization. Similarly, in the spinal cord, the FWHM volume decreases with higher frequencies; however, unlike in the thalamus, off-site focalization occurs, so the beam does not reach the spinal target. Figure 4C shows the relationship between frequency and beam overlap, where beam overlap indicates how much the FWHM overlaps with the target. The beam overlap is greater than 50% for the thalamus at all three frequencies, whereas it is near zero for the spinal cord.

The standard deviation (SD) of maximum sound intensity in the thalamus was 0.064, 0.100, and 0.141 at 300, 500, and 700 kHz, respectively, with corresponding coefficients of variation (CV) of 11%, 14%, and 8%. This indicates that targeting the spinal cord is more sensitive to small variability of transducer positioning than targeting the thalamus.

*3.2. Optimizing SEFT Geometry for tsFUS*

Optimal transducer size for spinal cord stimulation was investigated. The analysis was performed with constant stimulation (30 kPa) and frequency (500 kHz). The radius of curvature (*C*) of the transducer was varied from 60 mm to 85 mm, and the aperture width (*A*) from 35 mm to 100 mm, and a total of 55 SEFTs of different sizes, each of which did not overlap with the human body model, were simulated.

Figure 5A shows the results of the SEFT geometry analysis for spinal cord stimulation. The horizontal axis represents beam overlap, the vertical axis indicates the maximum intensity at the target, and each point corresponds to a different SEFT geometry defined by the ratio *C*/*A*. The results show that a higher ratio increases on-site focality but reduces the maximum acoustic intensity and vice versa. Based on this property, a Pareto front analysis was performed to simultaneously maximize the intensity and increase the beam overlap. As a result, the optimal SEFT size was found to be 80 mm for *C* and 55 mm for *A*. Figure 5B compares the intensity distribution of the original SEFT, balanced SEFT, and beam maximizing SEFT. The balanced SEFT has a beam overlap increased by 20% compared to the original, while the intensity remains above the assumed minimum threshold. Maximum beam overlap can be achieved up to 27, but it is below the threshold limit.

*3.3. Effect of Sonication Acoustic Paths Along Spinal Levels*

Figure 6 shows the effect of varying transducer position along the spinal levels when sonication is conducted through the posterior intervertebral acoustic window and vertebral lamina. Table 2 presents the maximum acoustic intensity within each spinal target. Sonication through the posterior acoustic window yields higher intensities at the thoracic levels compared to the laminar approach (1.1-fold to 6.4-fold increase). In contrast, lamina-based sonication presents higher intensities in the cervical levels than the acoustic window (1.1-fold to 8.7-fold, median of 2.3). During acoustic window-based sonication, Figure 6A shows a widespread beam with hotspots in the tendon at the C4–C5 cervical levels. Conversely, the beam is less widespread at T6–T7 but increases at T10–T11, although in both cases, hotspots reach the spinal cord at thoracic levels. During lamina-based sonication, Figure 6B shows that the beam is less affected at cervical levels compared to thoracic levels. Overall, beam overlap within the spinal cord is limited in lamina-based sonication due to obstruction by surrounding vertebral structures.

*3.4. Transducer Misalignment Effect*

Figure 7 shows the changes in ultrasound propagation when the transducer is shifted vertically by 10 and 20 mm (both upward and downward) relative to the posterior intervertebral acoustic window between T10 and T11 and relative to the laminar position at C4 and T6. The optimal transducer size ($C$ = 80 mm and A = 55 mm) was used, along with a frequency of 500 kHz and a source amplitude of 30 kPa. The figure illustrates the impact of bone barriers on acoustic transmission resulting from small positional variations. The maximum intensity for each position is presented in Table 3. The intensity decreases by a factor of 3.4 to 17.4 when moving outside the acoustic window. In lamina-based sonication, it is possible to increase the intensity by utilizing the interlaminar space, as seen in both C4 and T6 conditions.

## 4. Discussion

We numerically evaluated the feasibility of using trans-spinal focused ultrasound (tsFUS) to reach and precisely target spinal cord areas in humans. Neuromodulation of the spinal cord holds significant therapeutic potential. While a few studies have demonstrated neuromodulatory effects of FUS on the spinal cord in rodent models,[13,28] evidence in humans remains limited. In a prior preliminary conference paper, we demonstrated the potential of tsFUS to reach the human spinal cord.[29] More recently, spinal cord neuromodulation in humans was observed during sonication through cervical vertebrae.[20] The present study offers a comprehensive computational investigation using high-resolution, anatomically accurate human models to evaluate critical parameters such as acoustic intensity and beam overlap. These results provide foundational insights into the feasibility and optimization of tsFUS for noninvasive spinal cord stimulation, and they extend the emerging body of experimental research for humans.

*Sonication on Spinal Cord*

We used a commercial transducer for neuromodulation in the human brain and applied it to the spinal cord (acoustic window at T10-T11) in section 3.1. The results showed that at all frequencies, the maximum intensity delivered by tsFUS to the spinal cord was higher than the maximum intensity delivered by tFUS to the thalamus (1.5 times – 3.7 times). This is because the skull greatly attenuates ultrasound during brain stimulation,[30] whereas acoustic windows enable near-bone-free transmission, but a small acoustic window imposes aperture constraints that degrade beam shape symmetry, broaden the focal spot, and increase sidelobes, particularly when the window is small relative to the commercial transducer aperture. Indeed, the beam overlap was higher in the thalamus (50–70%) than in the spinal cord (smaller than 4%).[10,31]

Therefore, we adapted the original SEFT transducer for the posterior intervertebral acoustic window to improve focus on the spinal cord. In section 3.2, a trade-off was observed between beam overlap (focal accuracy) and intensity. By increasing the curvature ratio (C/A) of the SEFT geometry, focality improves while the acoustic intensity decreases. Selecting a SEFTS geometry (C = 80 mm, A = 55 mm) increases the beam overlap by 20% (from 2% to 22%) while maintaining the minimum acoustic intensity threshold required to elicit a biological response. Furthermore, the maximum beam overlap achieved was 27%, although the sound pressure input of the transducer should be increased to meet the minimum assumed acoustic intensity threshold. However, it was still far from the beam overlap levels in the brain (up to 70%). In general, higher ratios of C/A produce greater beam overlap, although they may reduce acoustic intensity, indicating a trade-off between focality and power.

*Acoustic Paths*

In addition to the posterior intervertebral acoustic window, vertebral lamina-based sonication was considered because of its thinner and flatter structure, which reduces reflection, refraction, and attenuation. Both acoustic paths were evaluated at cervical and thoracic levels in section 3.3.

Acoustic window-based sonication produces varying intensity levels at different vertebral levels, with cervical targeting resulting in smaller intensities than thoracic targeting by a factor of 6.5. The narrower acoustic window in the cervical region can explain this difference. Overall, sonication from T5–T6 to T11–T12 exhibits relatively high acoustic intensities, indicating that this segment offers favorable anatomical conditions for the acoustic path. Notably, the high intensities at T6–T7 and T7–T8 may also be attributed to the short length of the ligaments and the proximity of the skin surface to the spinal cord. Similarly, the higher intensity at T10–T11 and T11–T12 may be due to the larger size of the acoustic window in these regions, which allows for more effective ultrasound transmission.

Sonication through the lamina was investigated at spinal levels C1 to T12. This method resulted in higher intraspinal acoustic intensities than the posterior acoustic window in cervical levels (C1–C7), indicating that targeting through the lamina may provide an effective alternative path in the upper spine. This difference is likely attributable to the smaller intervertebral acoustic window in the cervical levels. In contrast, in the thoracic spine (T1–T12), the intervertebral acoustic window consistently yielded greater intensities than the lamina-based approach. Although adjusting the beam on interlaminar gaps may produce sonication levels as through the acoustic window (see below *Transducer Positioning*). These results suggest that optimal stimulation sites may need to be selected based on region-specific vertebral morphology. Nonetheless, the spatial beam overlap remains limited due to acoustic distortion and blockage by the vertebral arch.

*Transducer Positioning*

The posterior intervertebral acoustic window varies significantly depending on the spinal level, affecting the effectiveness and requirements for correct positioning of the transducer.[32] The results show that the misalignment of the transducer by 10 mm and 20 mm reduces the intensity by a factor of 3.4 to 17.4 when sonication is performed outside the posterior intervertebral window. This highlights the importance of prior identification of the acoustic window for guiding the transducer location.[33] Similarly, in lamina-based sonication at the C4 and T6 levels, the highest intensities were observed when the transducer was positioned at interlaminar spaces rather than directly centered over the lamina. These findings suggest that even for lamina-based approaches, transducer positioning has a significant impact on acoustic efficacy and should be optimized according to local anatomy. Given the critical role of dorsal root ganglia (DRG) in sensory transmission and their involvement in neuropathic pain syndromes, the transducer position to target DRG is investigated. DRG at T6–T7 could surpass the assumed neuromodulation threshold (Appendix A). However, the inability to consistently reach DRG at other levels, such as T10–T11, highlights the limitations of current transducer design.

*Limitations and Future Work.*

This is a numerical computational study that characterizes sonication in the spinal cord. Firstly, a single adult male anatomical model was used. However, individual variability can affect acoustic propagation and focality. Future research should incorporate diverse anatomical models to evaluate how these factors influence sonication outcomes and determine the applicability of tsFUS parameters across patient populations. Secondly, we adopted 0.77 W/cm² as a reference intensity for neuromodulation, based on prior neuromodulation studies.[10] However, no standardized sonication level for neuromodulation, particularly in the spinal cord, has been established.[34] This choice of reference intensity does not affect the observed tsFUS outcomes, as the input power of the transducer can be adjusted to achieve the required neuromodulation level. Thirdly, thermal modeling should be included in future work to ensure the safe selection of parameters,[35] especially when higher input power may be necessary due to low sonication levels reaching the spinal cord.[19] Fourthly, this study did not account for spinal posture changes, such as flexion, which may widen intervertebral and interlaminar spaces and improve bone-free acoustic paths. Finally, future studies can investigate other transducer configurations, such as phased-array systems, which offer the flexibility to account for vertebral anatomical complexity and enhance the precision of targeting dorsal root ganglia.

## 5. Conclusions

This study demonstrates the potential feasibility of trans-spinal focused ultrasound (tsFUS) for noninvasive, targeted neuromodulation of the human spinal cord through numerical analysis. By utilizing an intervertebral acoustic window for the thoracic vertebrae and optimizing the SEFT transducer geometry, we achieved better beam focus and sufficient intensity levels for neuromodulator that follow. We observed that sonication through the lamina is more effective than through the posterior intervertebral acoustic window for cervical vertebrae. Our results highlight the impact of anatomical factors and the need for precise transducer positioning to ensure effective sonication. Future research should address individual anatomical differences, develop tailored transducer designs, and verify these findings in experimental and clinical settings.

**Disclosure**

This study was supported by a JSPS Grant-in-Aid for Scientific Research, JSPS KAKENHI Grant 25K15887

**Conflict of Interest**

None

**Declaration of generative AI and AI-assisted technologies in the writing process**

During the preparation of this work, the authors used ChatGPT in order to check grammar. After using this tool, the authors reviewed and edited the content as needed and take full responsibility for the content of the published article.

**Appendix A**

*Stimulation to DRG*

The potential to stimulate the DRG was examined using the same transducer parameters as in section 3.3. DRG sonication was tested at thoracic levels T5–T6 through T12–L1, chosen based on their high acoustic intensities observed in Section 3.3. At each intervertebral level, the right-side DRG was designated as the target. To account for the transducer angle, the transducer was tilted by 5 and 10 degrees both vertically and horizontally at each target, as shown in Fig. A1-A, resulting in nine sonications for each target.

Figure A1 illustrates the distribution of the maximum intensity values within the DRG for each target region (n = 9). Stimulation at certain levels (e.g., T5-T6, T6–T7, T7–T8, and T12-L1) exceeded the reference intensity threshold of 0.77 W/cm² of biological response. Conversely, at levels such as T10–T11 and T11–T12, sonication was often reduced due to acoustic obstruction by vertebral structures, preventing the focus from adequately reaching the DRG. Figure A1 also presents representative acoustic intensity distributions. For example, at T6–T7 and T7–T8, the hotspot reached the DRG, whereas at T10–T11, the vertebrae blocked the beam, leading to insufficient targeting. These findings suggest that optimizing transducer angles may enable effective DRG stimulation at specific thoracic levels, although anatomical variability remains a significant challenge.

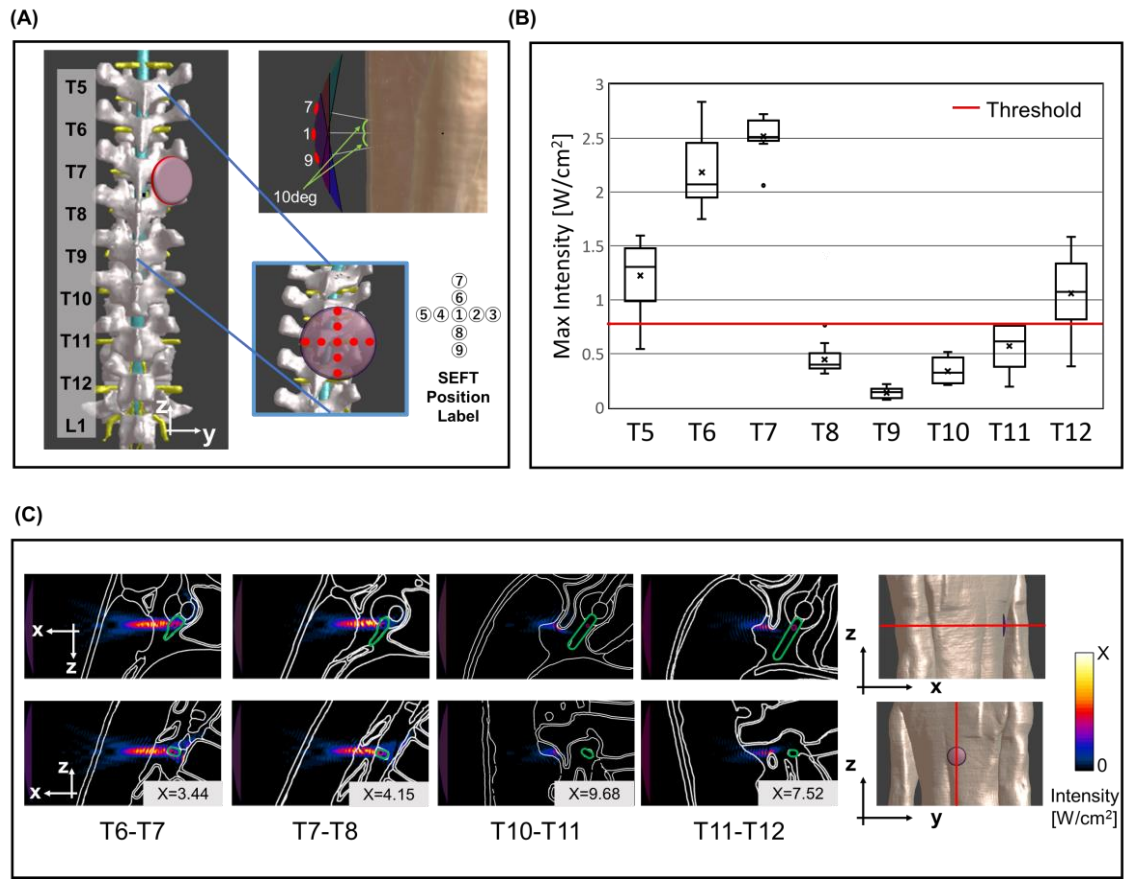

**Figure A1.** Sonication targeting the dorsal root ganglion (DRG). (A) Transducer placement relative to the vertebrae. (B) Distribution of maximum acoustic intensity values within DRG ($n = 9$). (C) Representative acoustic intensity distributions using SEFT configuration with C = 80 mm, and A = 55 mm. The target region is marked by a green line. The red line indicates an acoustic intensity of 0.77 W/cm², which corresponds to the biological response threshold.

**Tables and Figures**

**Table 1.** SEFT size

| Curvature radius (C) [mm] | Aperture width (A) [mm] |
|:---:|:---:|
| 60 | 35, 40…,60 |
| 65 | 40, 45…,75 |
| 70 | 45, 50…,90 |
| 75 | 50, 55…,100 |
| 80 | 55, 60…,100 |
| 85 | 55, 60…,100 |

**Table 2.** Maximum intensity in the spinal cord with intervertebral and lamina-based transducer placement from C1–C2 to T12–L1 (fixed input pressure of 30 kPa)

| SEFT position over the posterior acoustic window | Intensity [W/cm$^2$] | SEFT position over Lamina | Intensity [W/cm$^2$] |
|---|---|---|---|
| C1–C2 | 0.24 | C1 lamina | 0.50 |
| C2–C3 | 0.08 | C2 lamina | 0.31 |
| C3–C4 | 0.18 | C3 lamina | 0.40 |
| C4–C5 | 0.25 | C4 lamina | 0.56 |
| C5–C6 | 0.08 | C5 lamina | 0.27 |
| C6–C7 | 0.03 | C6 lamina | 0.26 |
| C7–T1 | 0.11 | C7 lamina | 0.12 |
| Average ± SD | 0.14 ± 0.08 | Average ± SD | 0.35 ± 0.14 |
| T1–T2 | 0.29 | T1 lamina | 0.26 |
| T2–T3 | 0.40 | T2 lamina | 0.31 |
| T3–T4 | 0.41 | T3 lamina | 0.17 |
| T4–T5 | 0.36 | T4 lamina | 0.25 |
| T5–T6 | 0.66 | T5 lamina | 0.32 |
| T6–T7 | 2.32 | T6 lamina | 0.64 |
| T7–T8 | 1.25 | T7 lamina | 0.25 |
| T8–T9 | 0.80 | T8 lamina | 0.24 |
| T9–T10 | 0.71 | T9 lamina | 0.33 |
| T10–T11 | 1.39 | T10 lamina | 0.25 |
| T11–T12 | 1.22 | T11 lamina | 0.19 |
| T12–L1 | 0.36 | T12 lamina | 0.19 |
| Average ± SD | 0.91 ± 0.61 | Average ± SD | 0.28 ± 0.12 |

**Table 3.** Intensity variation in the spinal cord due to misalignment from the central position, across different acoustic paths (C4 lamina, T6 lamina, T10–T11).

| Horizontal Shift [mm] | Intensity [W/cm$^2$] | | |
| --- | --- | --- | --- |
| | C4 lamina | T6 lamina | T10-T11 |
| -20 | 0.18 | 0.42 | 0.08 |
| -10 | 1.40 | 2.16 | 0.22 |
| 0 | 0.56 | 0.64 | 1.39 |
| 10 | 0.52 | 0.94 | 0.40 |
| 20 | 0.28 | 0.53 | 0.13 |

**Figure Legends**

**Figure 1. Anatomical targets and model setup for focused ultrasound stimulation (FUS).** (A) Posterior and lateral views of the human body showing the spinal cord and thalamus as targets, along with relevant overlying structures such as skin, vertebrae, skull, gray matter, and white matter. (B) Axial cross-section of the torso showing internal tissues, including the spinal cord and dorsal root ganglion (DRG) as sonication targets. (C) Vertebral column and surrounding structures. Sonication is conducted through the posterior intervertebral acoustic window and the vertebral lamina.

**Figure 2. Sonication setup and focal field estimation metrics.** (A) Diagram of the transducer geometry (spatially-extended focal target: SEFT) illustrating the aperture width (*A*) and curvature radius (*C*) for two cases. (B) Positioning of the SEFT transducer for spinal cord (top) and thalamus (bottom) targets. Red dots indicate stimulation points used to evaluate orientation-induced positioning errors. (C) Sonication of spatial intensity distribution in a cross-section of the spinal cord. The bottom image shows the full-width at half-maximum (FWHM) volume in red. Beam overlap is the percentage of the FWHM volume that is inside the target (green lines).

**Figure 3. Acoustic distributions at multiple frequencies for the spinal cord (through the acoustic window at T10-T11) and thalamus targets**. Intensity distributions for the spinal cord using the SEFT transducer (C = 85, A = 80) at 300 kHz (A, D, G, J), 500 kHz (B, E, H, K), and 700 kHz (C, F, I, L). Both targets are indicated by green lines.

**Figure 4. Comparison of acoustic metrics for thalamus and spinal cord at T10-T11 (n = 9).** (A) Peak spatial intensity. The red line indicates the acoustic intensity of 0.77 W/cm² at which the biological response was confirmed during tFUS at a 500 kHz stimulus. (B) Full-width at half-maximum (FWHM) volume. (C) Beam overlap percentages.

**Figure 5. (A) Relationship between acoustic peak intensity and beam overlap in the spinal cord for various SEFT transducer sizes, characterized by curvature/aperture (C/A) ratios.** Each point represents a different SEFT configuration. The red line indicates the acoustic intensity of 0.77 W/cm², corresponding to the threshold for biological response. Black-circled points denote the Pareto front. The best balanced SEFT size, based on Pareto front analysis, is identified as C = 80 mm and A = 55 mm. (B) Acoustic intensity distribution of the original SEFT, the balanced SEFT for spinal cord stimulation, and the SEFT achieving maximum beam overlap. The green frame indicates the target region (top row: transverse plane; bottom row: sagittal plane).

**Figure 6. Effect of acoustic path and transducer position along the spinal levels.** (A) Sonication through the posterior acoustic window. The left panel illustrates transducer placements in relation to the vertebral column from C1–C2 to T11–T12. The right plane shows acoustic intensity distributions. (B) Sonication through the vertebral lamina. The left panel illustrates transducer placements in relation to the vertebral column at the C4 and T6 levels. The right plane shows acoustic intensity distributions. The green frame indicates the target region.

**Figure 7. Impact of transducer misalignment relative to vertebral lamina (C4 lamina and T6 lamina) and posterior the acoustic window (T10–T11)** using the balanced SEFT configuration (C = 80 mm, A = 55 mm). (A) Transducer positions relative to the vertebral column at the C4 and T6 levels (lamina approach) and at the T10–T11 level (posterior acoustic window). (B) Acoustic intensity distributions. For each transducer position, acoustic fields are shown in two orthogonal planes: one along the beam axis and one perpendicular to it. The green frame designates the target region.

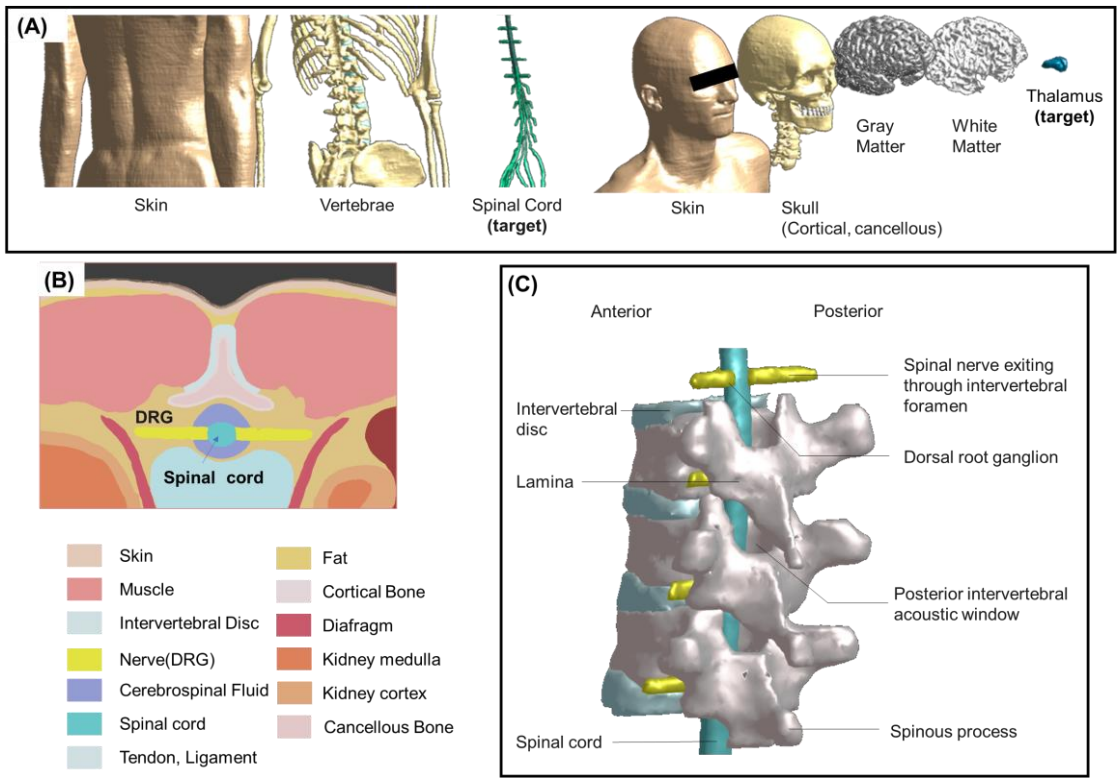

**Figure 1**

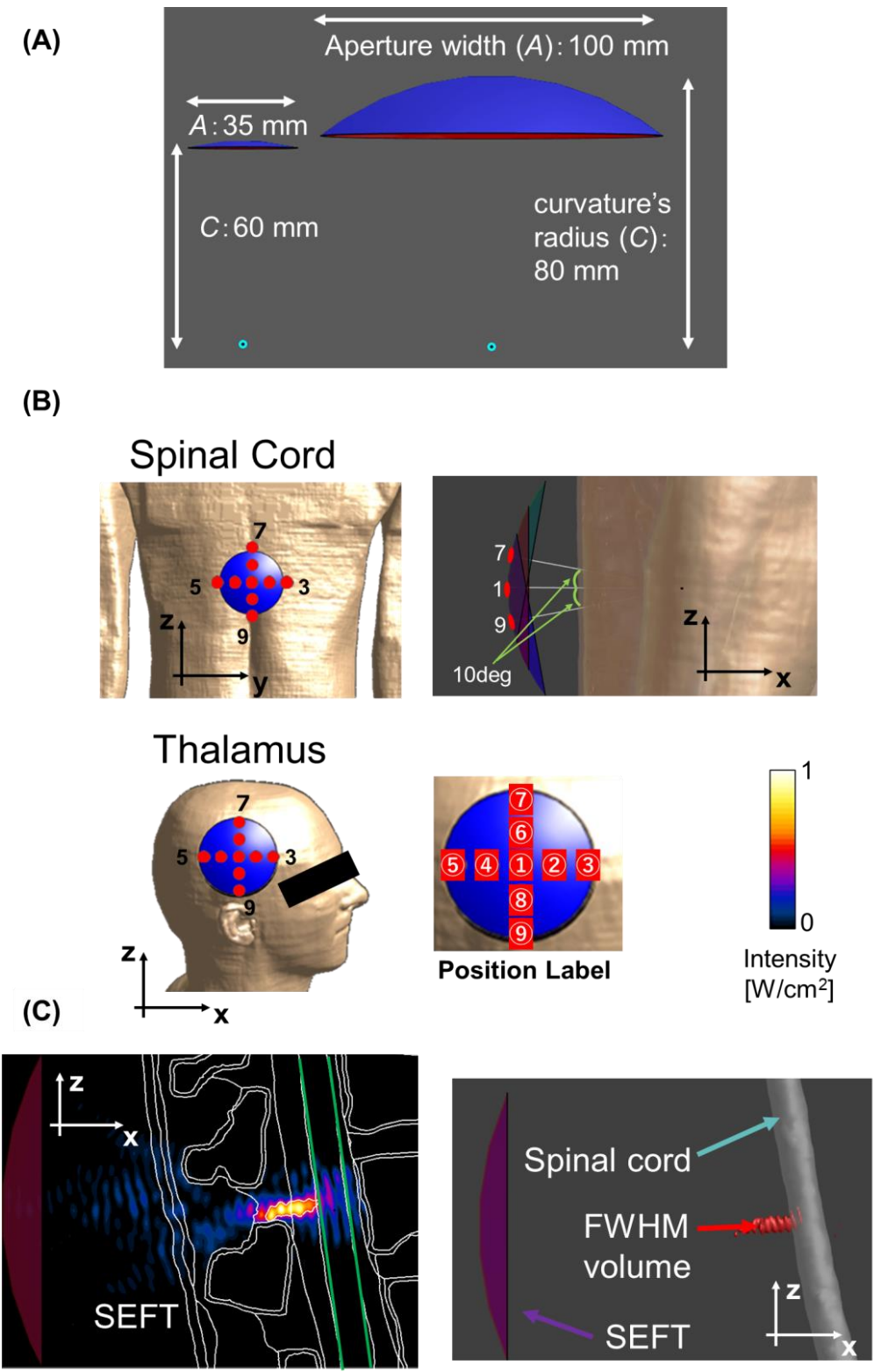

**Figure 2**

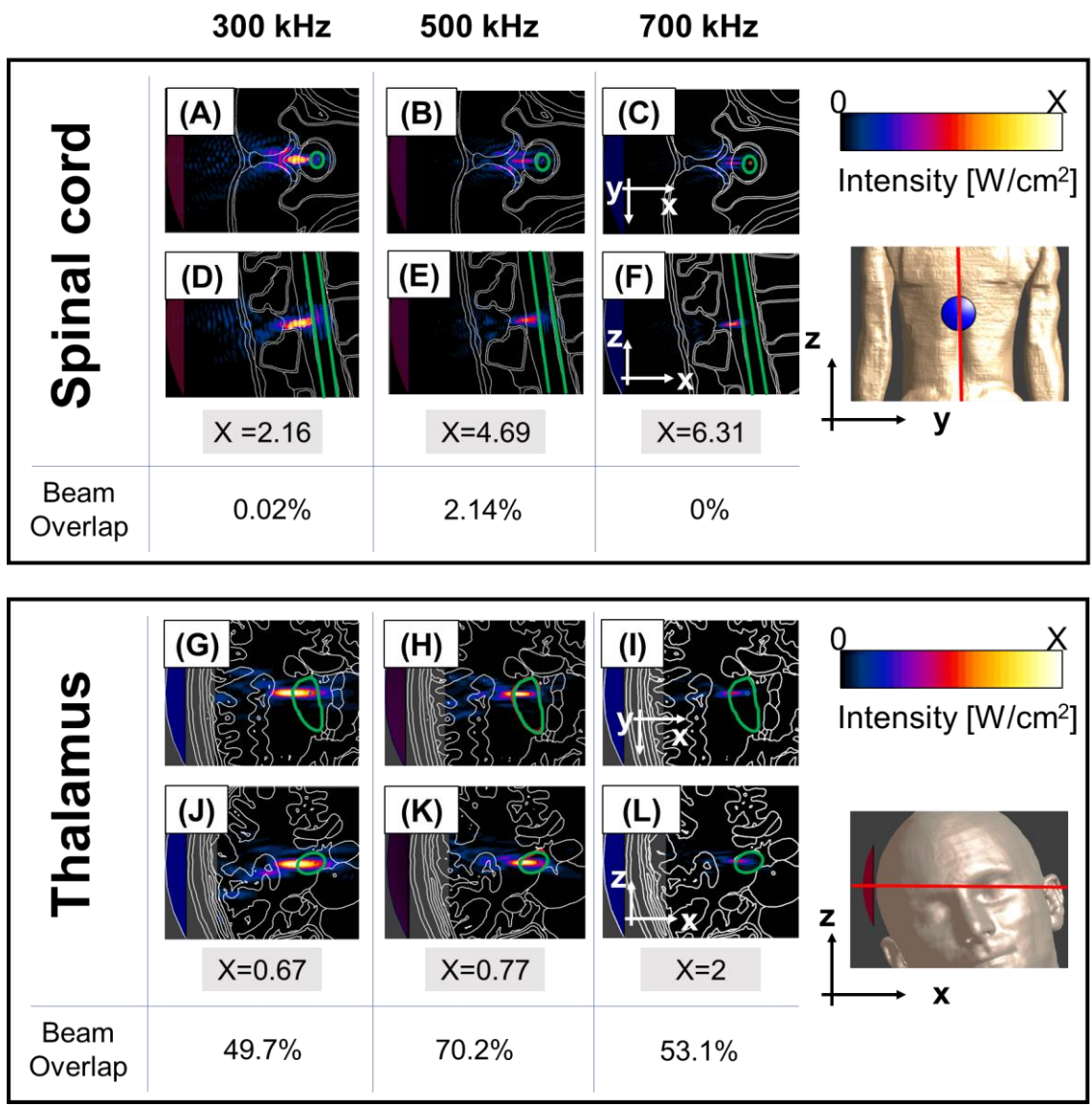

**Figure 3**

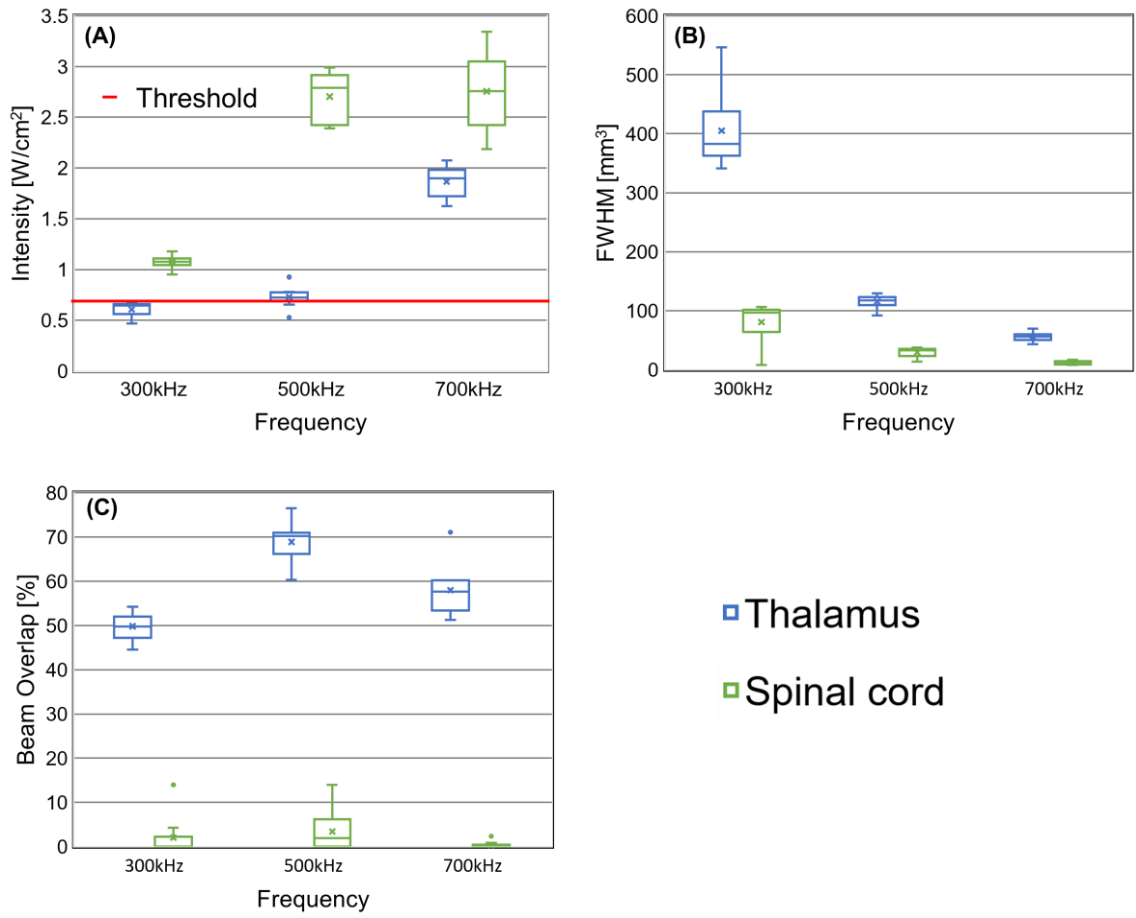

**Figure 4**

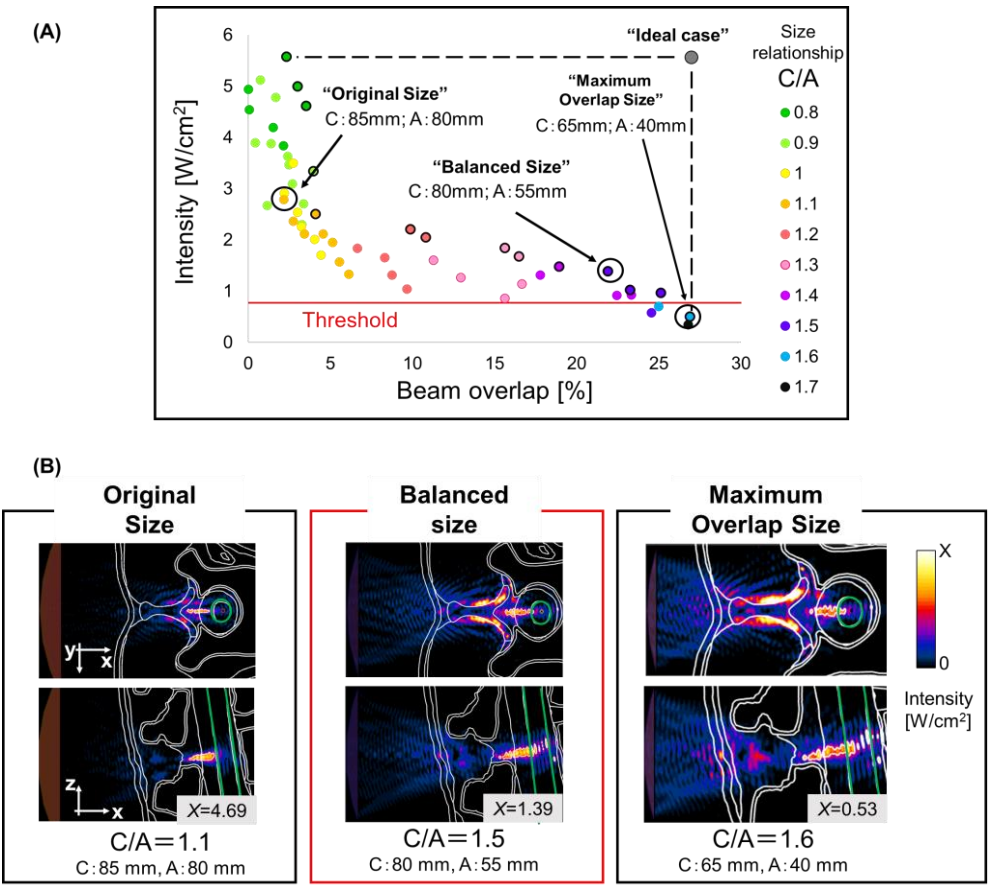

**Figure 5**

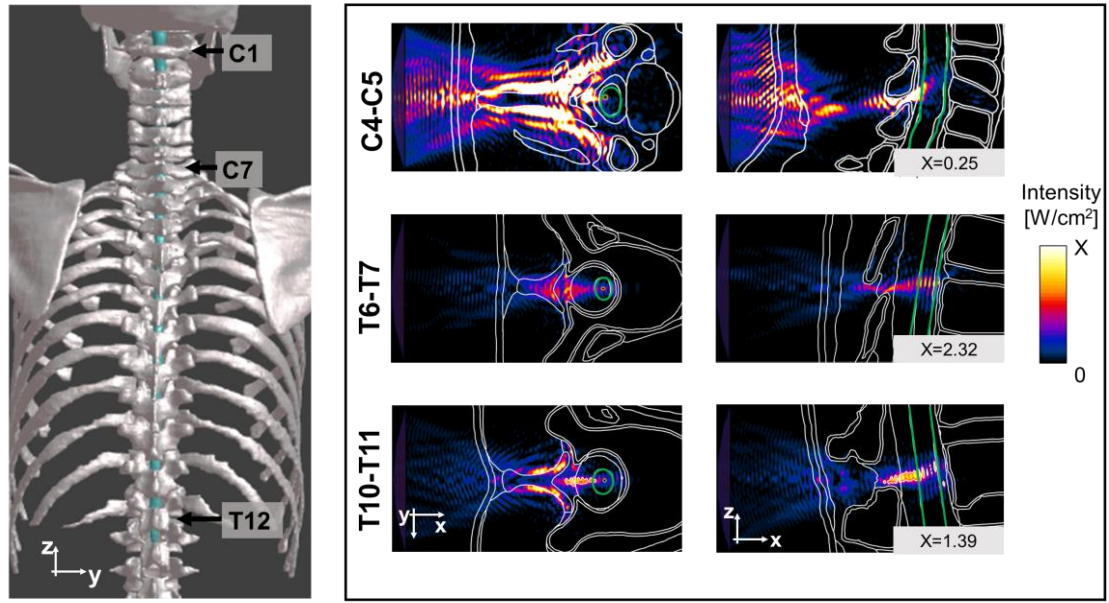
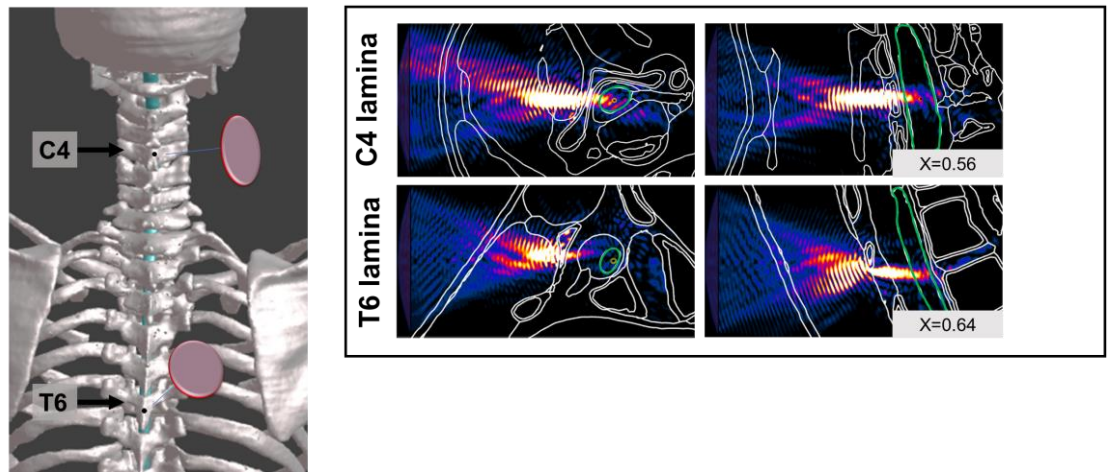

**Figure 6**

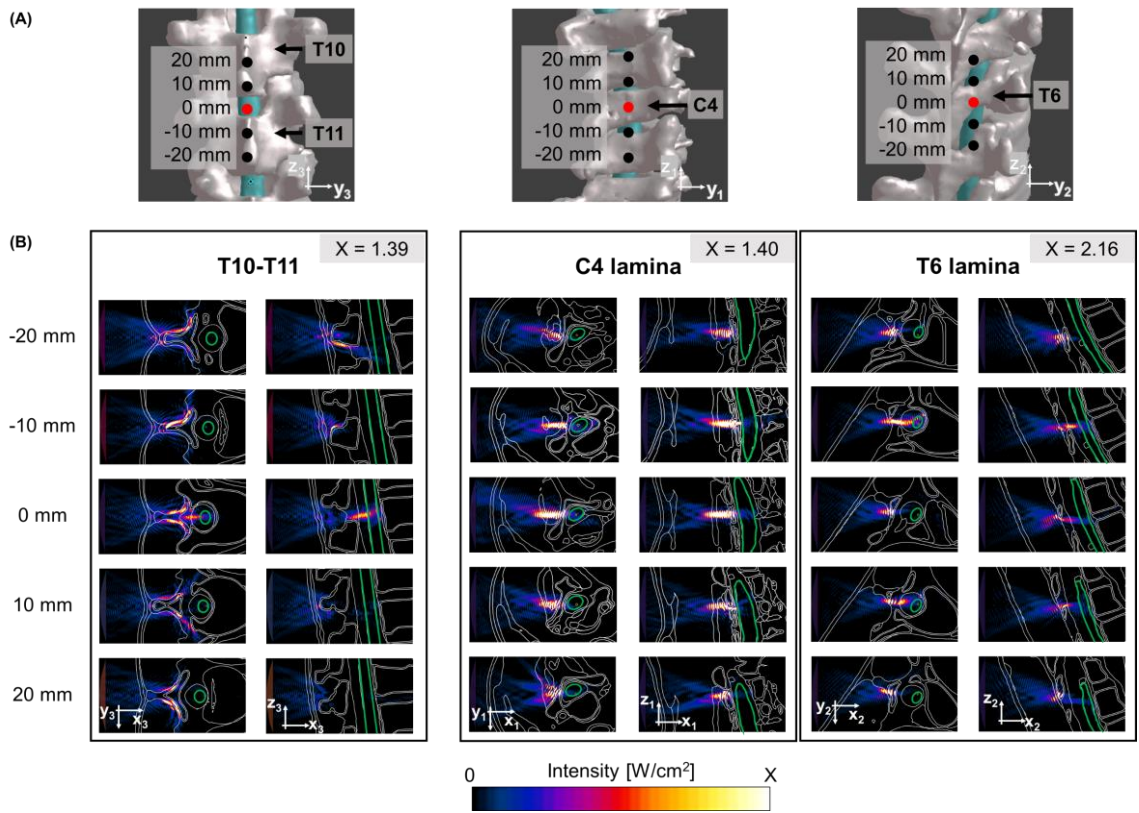

**Figure 7**